\documentclass[fleqn,10pt]{article}
\newcommand{\onlinecite}[1]{\hspace{-1 ex} \nocite{#1}\citenum{#1}}

\RequirePackage[utf8]{inputenc}
\RequirePackage[english]{babel}

\RequirePackage{ifthen}
\RequirePackage{calc}
\AtEndOfClass{\RequirePackage{microtype}}
\RequirePackage{times}      
\RequirePackage{ifpdf}

\RequirePackage{amsmath,amsfonts,amssymb}
\RequirePackage{graphicx,xcolor}
\RequirePackage{booktabs}

\RequirePackage{authblk}
\RequirePackage[left=2cm,%
                right=2cm,%
                top=2.25cm,%
                bottom=2.25cm,%
                headheight=12pt,%
                letterpaper]{geometry}%
                
\RequirePackage[labelfont={bf,sf},%
                labelsep=period,%
                justification=raggedright]{caption}

\RequirePackage[colorlinks=true, allcolors=blue]{hyperref}

\RequirePackage[superscript,biblabel]{cite}
\RequirePackage{fancyhdr}  
\RequirePackage{lastpage}  
\RequirePackage[explicit]{titlesec}
\titleformat{\section}
  {\color{color1}\large\sffamily\bfseries}
  {\thesection}
  {0.5em}
  {#1}
  []
\titleformat{name=\section,numberless}
  {\color{color1}\large\sffamily\bfseries}
  {}
  {0em}
  {#1}
  []  
\titleformat{\subsection}
  {\sffamily\bfseries}
  {\thesubsection}
  {0.5em}
  {#1}
  []
\titleformat{\subsubsection}
  {\sffamily\small\bfseries\itshape}
  {\thesubsubsection}
  {0.5em}
  {#1}
  []    
\titleformat{\paragraph}[runin]
  {\sffamily\small\bfseries}
  {}
  {0em}
  {#1} 
\titlespacing*{\section}{0pc}{3ex \@plus4pt \@minus3pt}{5pt}
\titlespacing*{\subsection}{0pc}{2.5ex \@plus3pt \@minus2pt}{0pt}
\titlespacing*{\subsubsection}{0pc}{2ex \@plus2.5pt \@minus1.5pt}{0pt}
\titlespacing*{\paragraph}{0pc}{1.5ex \@plus2pt \@minus1pt}{10pt}

\usepackage{titletoc}
\contentsmargin{0cm}
\titlecontents{section}[\tocsep]
  {\addvspace{4pt}\small\bfseries\sffamily}
  {\contentslabel[\thecontentslabel]{\tocsep}}
  {}
  {\hfill\thecontentspage}
  []
\titlecontents{subsection}[\tocsep]
  {\addvspace{2pt}\small\sffamily}
  {\contentslabel[\thecontentslabel]{\tocsep}}
  {}
  {\ \titlerule*[.5pc]{.}\ \thecontentspage}
  []
\titlecontents*{subsubsection}[\tocsep]
  {\footnotesize\sffamily}
  {}
  {}
  {}
  [\ \textbullet\ ]  
  
\RequirePackage{enumitem}

\renewcommand{\@biblabel}[1]{\bfseries\color{color1}#1.}

\newcommand{\keywords}[1]{\def\@keywords{#1}}

\def\xabstract{abstract}
\long\def\abstract#1\end#2{\def\two{#2}\ifx\two\xabstract 
\long\gdef\theabstract{\ignorespaces#1}
\def\go{\end{abstract}}\else
\typeout{^^J^^J PLEASE DO NOT USE ANY \string\begin\space \string\end^^J
COMMANDS WITHIN ABSTRACT^^J^^J}#1\end{#2}
\gdef\theabstract{\vskip12pt BADLY FORMED ABSTRACT: PLEASE DO
NOT USE {\tt\string\begin...\string\end} COMMANDS WITHIN
THE ABSTRACT\vskip12pt}\let\go\relax\fi
\go}

\renewcommand{\@maketitle}{%
{%
\thispagestyle{empty}%
\vskip-36pt%
{\raggedright\sffamily\bfseries\fontsize{20}{25}\selectfont \@title\par}%
\vskip10pt
{\raggedright\sffamily\fontsize{12}{16}\selectfont  \@author\par}
\vskip18pt%
{%
\noindent
{\parbox{\dimexpr\linewidth-2\fboxsep\relax}{\color{color1}\large\sffamily\textbf{ABSTRACT}}}
}%
\vskip10pt
{%
\noindent
\colorbox{color2}{%
\parbox{\dimexpr\linewidth-2\fboxsep\relax}{%
\sffamily\small\textbf\\\theabstract
}%
}%
}%
\vskip25pt%
}%
}%
\definecolor{color1}{RGB}{0,0,0} 
\definecolor{color2}{gray}{1} 

\newlength{\tocsep} 
\usepackage{lipsum} 
\let\oldbibliography\thebibliography
\renewcommand{\thebibliography}[1]{%
\addcontentsline{toc}{section}{\hspace*{-\tocsep}\refname}%
\oldbibliography{#1}%
\setlength\itemsep{0pt}%
}

\usepackage{amsmath,amssymb,amsthm}
\usepackage{graphicx}
\newcommand{\be}{\begin{equation}}
\newcommand{\ee}{\end{equation}}
\newcommand{\ba}{\begin{eqnarray}}
\newcommand{\ea}{\end{eqnarray}}
\newcommand{\ban}{\begin{eqnarray*}}
\newcommand{\ean}{\end{eqnarray*}}

\newcommand{\ket}[1]{\mbox{$ | #1 \rangle $}}
\newcommand{\bra}[1]{\mbox{$ \langle #1 | $}}

\begin{document}
\title{Entanglement convertibility by sweeping through the quantum phases of the alternating bonds \textit{XXZ} chain}

\author[1,2,*]{Yu-Chin Tzeng}
\author[1,*]{Li Dai}
\author[1]{M.-C. Chung}
\author[3,4]{Luigi Amico}
\author[4,5,6,7]{Leong-Chuan Kwek}
\affil[1]{Department of Physics, National Chung Hsing University, Taichung 40227, Taiwan}
\affil[2]{Center of Theoretical Sciences, National Taiwan University, Taipei 10617, Taiwan}
\affil[3]{CNR-MATIS-IMM \& Dipartimento di Fisica e Astronomia, Universit\'a di Catania, \& INFN-Laboratori Nazionali del Sud, Via S. Sofia 64, 95127 Catania (Italy)}
\affil[4]{Centre for Quantum Technologies, National University of Singapore, 3 Science Drive 2, Singapore 117543, Singapore}
\affil[5]{Institute of Advanced Studies, Nanyang Technological University, 60 Nanyang View, Singapore 639673, Singapore}
\affil[6]{National Institute of Education, Nanyang Technological University, 1 Nanyang Walk, Singapore 637616, Singapore}
\affil[7]{MajuLab, CNRS-UNS-NUS-NTU International Joint Research Unit, UMI 3654, Singapore}
\affil[*]{These authors contributed equally to this work.}

\begin{abstract}
\textbf{We study the entanglement structure and the topological edge states of the ground state of the spin-1/2 \textit{XXZ} model with bond alternation. We employ parity-density matrix renormalization group with periodic boundary conditions. The finite-size scaling of R\'enyi entropies $S_2$ and $S_\infty$ are used to construct the phase diagram of the system. The phase diagram displays three possible phases: Haldane type (an example of symmetry protected topological ordered phases), Classical Dimer and N\'eel phases, the latter bounded by two continuous quantum phase transitions. The entanglement and non-locality in the ground state are studied and quantified by the entanglement convertibility. We found that, at small spatial scales, the ground state is not convertible within the topological Haldane dimer phase. The phenomenology we observe can be described in terms of correlations between edge states. We found that the entanglement spectrum also exhibits a distinctive response in the topological phase: the effective rank of the reduced density matrix displays a specifically large ``susceptibility'' in the topological phase. These findings support the idea that although the topological order in the ground state cannot be detected by local inspection, the ground state response at local scale can tell the topological phases apart from the non-topological phases.}
\end{abstract} 

\date{\today}
\flushbottom
\maketitle

\let\thefootnote\relax\footnote{Correspondence and requests for materials should be addressed to Y.C.T. (email: \texttt{d102054002@mail.nchu.edu.tw}) on condensed matter and L.D. (email: \texttt{lidaisgp@gmail.com}) on quantum information.}

\section*{Introduction}\label{Sec-1-intro}
\Large{\textbf{M}}\normalsize{any-body} quantum states are generically entangled. Consequently, considerable efforts have been made to understand  the physical implications of 
this simple fact\cite{etg-many-body-RMP,etg-RMP}. 
The issue is very challenging since  the entanglement found in many-body quantum states is highly multipartite, and it is distributed in a complex form\cite{multipartite_Briegel,PhysRevA_85_022301} with complicated structure~\cite{Wen}. Nevertheless, important results have been obtained for characterizing quantum phases of matter, or the phases of ground states, in terms of their entanglement.~\cite{Amico2002} Aside from disordered and ordered phases characterized by a non-vanishing (local) order parameter, quantum phases with more subtle order have since been found. This is the case of spin liquids and topologically ordered ground states\cite{yan2011spin}. These  phases,  and the phase transitions between them, cannot be characterized
within the mechanism of Landau symmetry breaking. Yet, these phases and their phase transitions exhibit patterns of long range entanglement that implies that a correlation, although tenuous, can be  of global nature (respect to a local scale implied by the bare interaction)~\cite{wen2004quantum}.  

\twocolumn
In this paper, we analyze the  entanglement structure of a quantum state through  the entanglement convertibility.
The latter is concerned with the conversion between quantum states through the Local Operations and Classical Communication (LOCC). This convertibility can be  used to characterize the quantum entanglement of the states. The rationale for that is fairly simple:
the convertibility properties of the state define equivalent class for entangled states, and in each class the states of multiple copies can be transformed into one another through LOCC which does not change the entanglement.
For bipartite pure states, meaning that the system can be divided into two subsystems A and B, the best ratio of $\frac{M'}{M}$ if $M$ copies of $\ket{\psi_{AB}}$ are converted into $M'$ copies of $\ket{\psi'_{AB}}$ (entanglement of formation and entanglement of distillation) is provided by $\frac{S_{\textrm{v}}(\rho_A)}{S_{\textrm{v}}(\rho'_A)}$, where $S_{\textrm{v}}(\rho_A)=-\textrm{Tr}(\rho_{A}\ln\rho_{A})$ is the von Neuman entropy of  $\rho_A=\textrm{Tr}_B(\ket{\psi_{AB}}\bra{\psi_{AB}})$ and  $\rho'_A$ is the reduced state of the subsystem A for $\ket{\psi'_{AB}}$~\cite{etg-entropy-conversion-ratio,Nielsen}. 
However, the scenario is very different when only a single copy of the state is available for converting  into another single copy of the target state through LOCC. Interestingly, in this case, it is not always possible to convert a state exactly into another state with the same or lower entanglement using only LOCC. To quantify the single copy entanglement conversion, we need to go beyond the informations provided by the von Neuman entropy, and a more complete knowledge of the eigenvalues of the reduced density operator is adduced through the Entanglement Spectrum (ES)~\cite{Haldane-ES} or, equivalently, through the R\'enyi entropies which provides a re-parametrization of the ES:
\ba\label{Renyi-entropy}
S_{\alpha}(\rho_A)=\frac{1}{1-\alpha}\ln(\textrm{Tr}\rho_A^{\alpha}),
\ea
where $\alpha\geq 0$. The case $\alpha=1$ corresponds to the von Neumann entropy. To the best of our knowledge, the single copy entanglement conversion cannot be expressed as a simple condition on the ES if a  {\it catalyst} is not involved in the process of  the conversion.~\cite{catalyst-1,catalyst-2} The catalyst here is a bipartite state that participates in the conversion process but remains intact after the conversion is done. In such a case, the necessary and sufficient condition for the convertibility of the state $\ket{\psi_{AB}}$ into $\ket{\psi_{AB}'}$ is that $S_{\alpha}(\rho_A)\geq S_{\alpha}(\rho_A')$ for all $\alpha\geq 0$.~\cite{catalyst-1,catalyst-2} Whether or not the catalyst is necessary in the conversion depends on the following majorization condition. Suppose the eigenvalues of $\rho_A$ are $(\omega_0,\omega_1,\omega_2,\cdots )\equiv\boldsymbol{\omega}$, where $\omega_0\geq\omega_1\geq\omega_2\geq\cdots$ is arranged in a non-increasing order. Similarly, the eigenvalues of $\rho'_{A}$ are $(\omega'_0,\omega'_1,\omega'_2,\cdots )\equiv\boldsymbol{\omega'}$. If $\sum_{j=0}^{k}\omega_j\leq\sum_{j=0}^{k}\omega'_j$ for all $k$, we say that $\boldsymbol{\omega}$ is majorized by $\boldsymbol{\omega'}$.~\cite{majorization}. The conversion from $\ket{\psi_{AB}}$ to $\ket{\psi_{AB}'}$ needs a catalyst if $\boldsymbol{\omega}$ is not majorized by $\boldsymbol{\omega'}$. Otherwise, the catalyst is not necessary.

\begin{figure*}[t]
\centering
\includegraphics[width=6in]{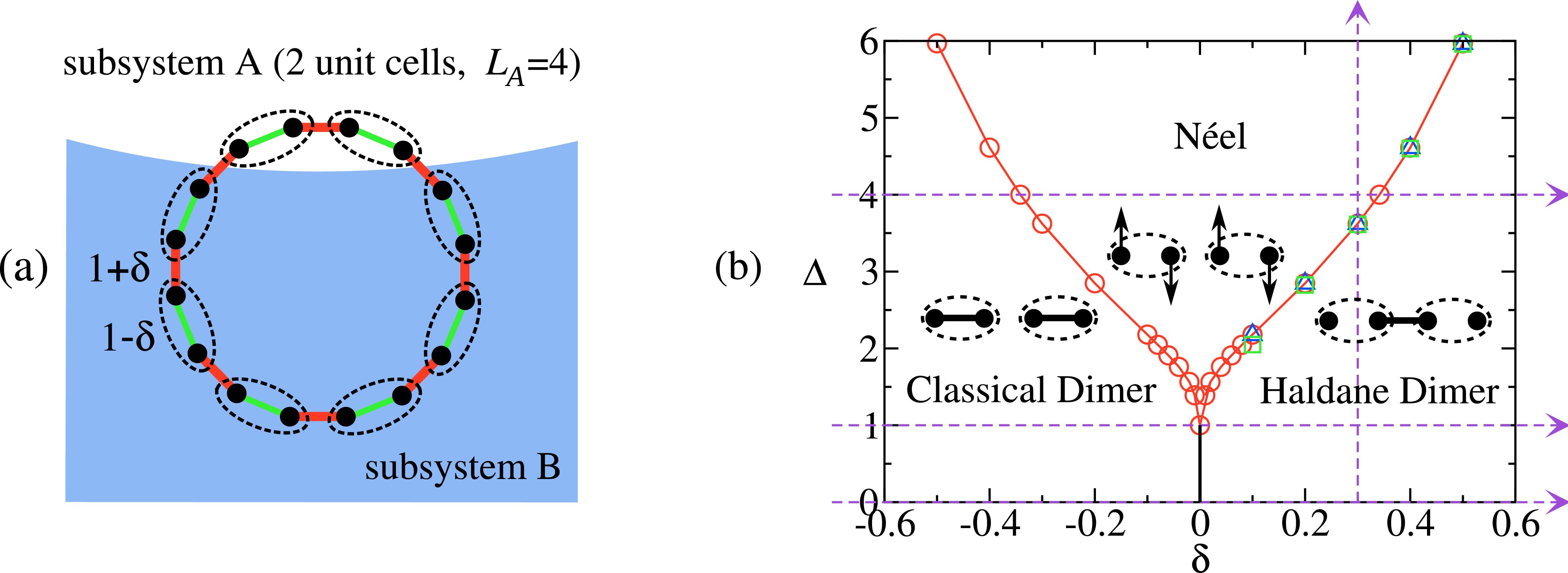} 
\caption{(a) The bond alternating spin-1/2 \textit{XXZ} model with length $N$. The dashed ellipses define $N/2$ unit cells of the system. The coupling strength of spins in the unit cell is $1-\delta$ (green lines), and between nearest units is $1+\delta$ (red lines). The subsystem \textit{A} contains complete unit cells. (b) The quantum phase diagram of the model is determined by using the finite-size scaling of R\'enyi entropies $S_2$ ($\bigtriangleup$), $S_\infty$ ($\Box$) and the second derivative of ground state energy ($\bigcirc$). The four dashed lines denote the routes that will be swept along.}\label{fig:BAxxz}
\end{figure*}

In this paper, we elaborate on the idea that important informations on  the entanglement structure encoded in the quantum phases of matter is possible through the study of the {\it response of the ES} to a perturbation of the ground state.  Such a response can be rendered into specific  convertibility properties of the ground state.\cite{DLC-XY,DLC-XY-2}
We study the Differential Local Convertibility (DLC), which is defined as  the convertibility between two ground states $\ket{\psi(g)}$ and $\ket{\psi(g+\epsilon)}$, corresponding to two Hamiltonians described by parameters $g$ and $g+\epsilon$ (where $\epsilon$ is an infinitesimal value).  Such an approach has been applied to quantum phases with meaningful order parameter~\cite{DLC-XY,DLC-XY-2,DLC-Ising}  and  topological phases of  two dimensional\cite{DLC-toric-code-1,DLC-toric-code-2} and one dimensional (1d)~\cite{DLC-Cluster-XXZD} spin models (the latter are examples of  class of spin liquids known as Symmetry-Protected Topological  phases~\cite{SPT-1,SPT-2,AKLT1987}; see Ref.~\onlinecite{PhysRevA.84.022304,0295-5075-95-5-50001,Son2011} for an example raised in quantum technology recently).  From these studies, we found that while disordered and symmetry broken phases are generically convertible, the topological phases are not and they violate the DLC property.   
Such property ultimately depends on the interplay between the spin-spin correlation length and the size of the partition:  When the subsystem size is  smaller than the correlation length between  the edges states  (found at the interface between $A$ and $B$) then the edge states can dominate this convertibility. In this case,  the ground state are not DLC. 
In the opposite limit the states are DLC. Such picture is also confirmed by the recent study on the Kitaev chain with a quenched chemical potential~\cite{Dai2015}.
Interestingly, in the cases of large partitions, it was found~\cite{DLC-XXZ-non-universal} that DLC can detect specific symmetries of the system.

Most of the current studies analyzed the quantum phase transition across two quantum phases. In this paper, we consider a more complex phase diagram with multiple quantum phases. Specifically,  we sweep trough two consecutive quantum phase transitions, delinating three quantum phases, one of which being an Symmetry Protected Topological Order (SPTO) phase.  We believe that this is an interesting case because 
the occurrence of quantum phase transitions, generally implies specific  constraints on  the behavior of the R\'enyi entropies in the phase diagram of the system. Therefore, we expect that 
the presence of multiple quantum phase transitions, can led to an ``interference''  effect, in particular, the slopes of the R\'enyi entropies and ultimately the DLC of the quantum phases.  
To this end, we study the spin-1/2 \textit{XXZ} chain with bond alternation whose phase diagram is provided in Fig. \ref{fig:BAxxz}(b). The techniques we employ are the parity Density Matrix Renormalization Group (pDMRG) with periodic boundary conditions~\cite{Tzeng2012} and the correlation function matrix~\cite{CFM-Peschel,free-fermion-CFM} for non-interacting fermions. We study the DLC through an analysis of the R\'enyi entropies. We study the majorization condition as well. The question we wish to study is whether we can characterize the different phases of the phase diagram in terms of the need for a catalyst. Incidentally, we note that just like the model studied in Ref.~\onlinecite{DLC-XXZ-non-universal}, our model can display an SU(2) symmetry non-critical point within the SPTO phase. We analyze the DLC at that point to leverage the analysis carried out in Ref.~\onlinecite{DLC-XXZ-non-universal}.

The paper is organized as follows. In following section, the model of spin-1/2 \textit{XXZ} chain with bond alternation is introduced and the calculation method is summarized. In the section ``Results'', the results on DLC are presented. For the topological phase, the mechanism of edge states recombination is used to interpret the inconvertibility. For the N\'eel phase, the three-phase mechanism is used to interpret inconvertibility. Finally, the conclusion is given. In the section ``Methods'', the correlation function matrix formula is discussed in detail, and the phase diagram is determined by the energy derivatives as well as the R\'enyi entropy $S_2$ and $S_\infty$.

\section*{Model Hamiltonian}\label{Sec-model-method}
The Hamiltonians of the 1d spin-$1/2$ alternating \textit{XXZ} model reads
\be\label{BA-XXZ-H}
  H=\sum_{n=1}^N[1+(-1)^n\delta ]
  (\sigma_n^x\sigma_{n+1}^x+\sigma_n^y\sigma_{n+1}^y+\Delta\sigma_n^z\sigma_{n+1}^z).
\ee
Here, $\vec\sigma_n$ are the Pauli matrices on the $n$th site of the chain with $N$ spins. Periodic boundary conditions, $\vec\sigma_{N+1}\to\vec\sigma_1$, are applied.  $\Delta$ is the strength of the Ising-type anisotropy which originates from the spin-orbit interaction in magnetic materials. $\delta$ is the bond alternation describing the dimerization by the spin-Peierls instability. The ground-state phase diagram is displayed in Fig.~\ref{fig:BAxxz}(b). We remark that we have drawn it using the finite-size scaling of R\'enyi entropies and the second derivative of ground state energy. Other methods such as the von Neumann entropy and the ground-state fidelity have been used to obtain a schematic phase diagram.~\cite{Tian2013} The Hamiltonian can be experimentally realized in ion traps and optical lattices, where the bond alternation is achieved by fine tuning the intensity of the Raman laser beams.~\cite{expriment-optical-lattice-1,expriment-optical-lattice-2,expriment-ion-trap}

\begin{figure*}[t]
\centering
\includegraphics[width=6in]{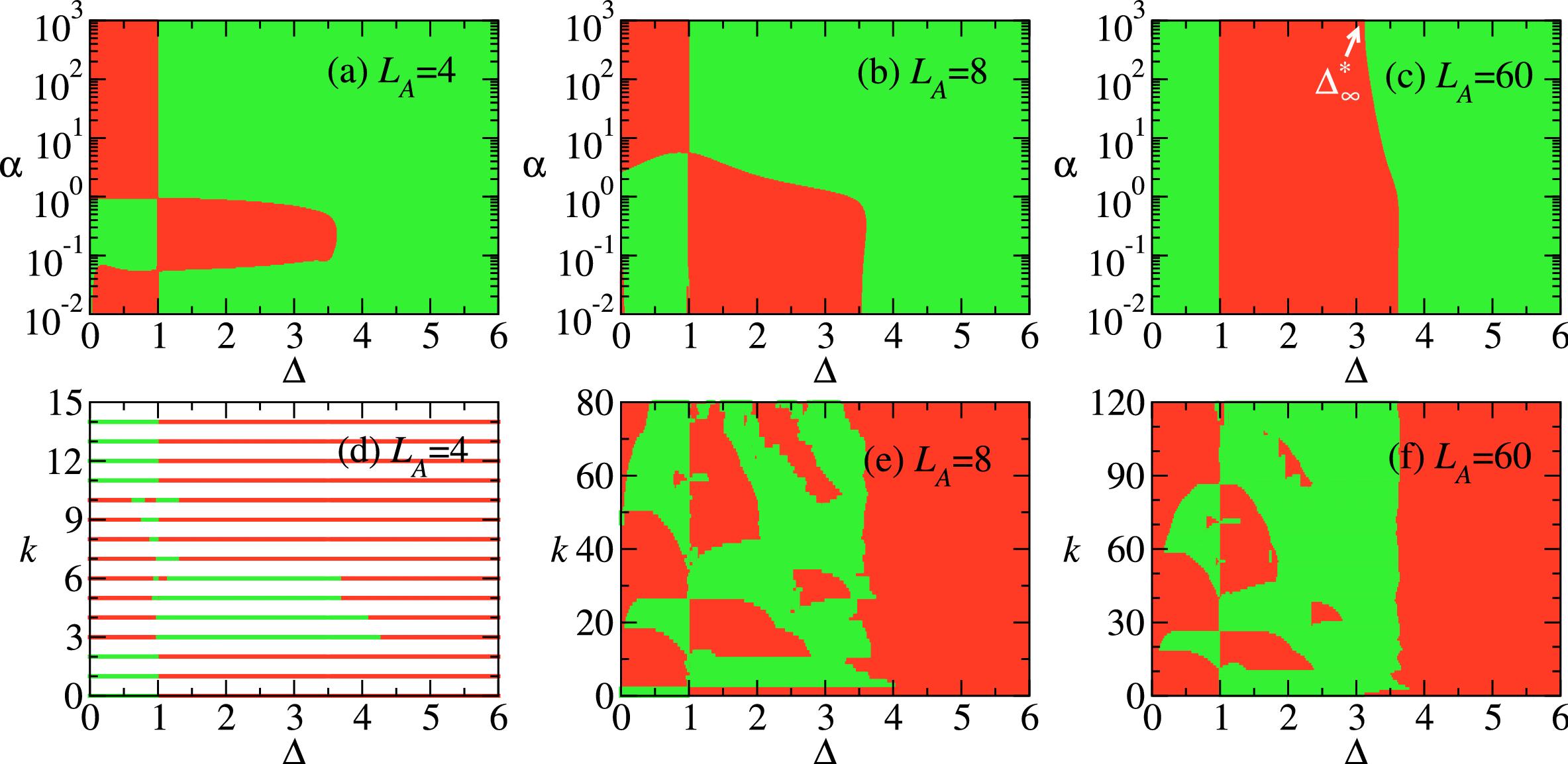}
\caption{\textbf{The differential local convertibility.} (a-c) The sign of $\frac{\partial S_{\alpha}}{\partial\Delta}$ is plotted on the $\alpha$-$\Delta$ plane,, where $S_{\alpha}$ is the R\'enyi entropy of the reduced state of subsystem A with size $L_{A}=4,8,60$ as marked in the graphs. (d-f) The corresponding plots of majorization $M(k)$ on the $k$-$\Delta$ plane. The red color denotes that $\frac{\partial S_{\alpha}}{\partial\Delta}>0$ ($M(k)>0$) and the green color indicates that $\frac{\partial S_{\alpha}}{\partial\Delta}<0$ ($M(k)<0$). The parameters are $N=120$, $\delta=0.3$, $\epsilon=5\times 10^{-3}$. In the DMRG calculation, $m=600$ states are kept, with the truncation error below $10^{-12}$.}\label{DLC-d0.3}
\end{figure*}

The isotropic limit of the model ($\Delta=1$) has been studied intensively.~\cite{Cross1979,Hida1992, Nakamura2002,HHHung2005,BAiso} By choosing the unit cell (site $2n-1$ and $2n$), as shown in Fig.~\ref{fig:BAxxz}(a), the ground state for $\delta>1$ (ferro-antiferromagnetic alternation) is numerically shown~\cite{Hida1992,HHHung2005} to approach the spin-1 Haldane system with a finite value of the non-local string order parameter. The nearest two spins in two different unit cells tend to form a dimer (singlet) or the so-called valence bond, and the unit cells approaches to a spin-1 chain, as described by Affleck, Kennedy, Lieb, and Tasaki.~\cite{AKLT1987} The ground state does not break the translational symmetry by translating a unit cell. On the other hand, a small dimerization $\delta>0$ breaks the translational symmetry (translating a lattice site) and opens a spin gap from the gapless Luttinger liquid state. The ground state becomes static dimers. Since it has been shown that there is no quantum phase transition between the dimer region ($0<\delta\leq 1$) and the Haldane region ($\delta>1$),~\cite{Hida1992} we refer to the phase as the Haldane-Dimer (HD) phase. On the other hand, for $\delta<0$, the unit cells tend to form dimers, and the ground state tends to be product over all unit cells. We therefore refer to the phase as the Classical Dimers (CD). Note that after choosing the unit cell, one can distinguish the HD from CD by studying the parity of the entanglement between subsystem A and B (in order to detect entanglement, the subsystem A should contain complete unit cells, \textit{i.e.} the size of the subsystem A, $L_A$ should be even).

In the absence of bond alternation $\delta=0$, the model can be solved  using the Bethe ansatz, and the ground-state phase diagram exhibits ferromagnetic ($\Delta\leq -1$), Luttinger liquid ($-1<\Delta\leq 1$), and the N\'eel ($\Delta >1$) phases.~\cite{QP1D} The phase boundaries for $\delta>0$ and $\delta<0$ are symmetric by translation of elementary lattice spacing. In the limit of $\Delta\to\infty$, the ground state is expected to manifest the antiferromagnetic N\'eel order ($\uparrow\downarrow\uparrow\downarrow$) for $\delta<1$ and double N\'eel order ($\uparrow\uparrow\downarrow\downarrow$) for $\delta>1$, separated by the decoupled line $\delta=1$.
Both N\'eel states are nearly trivial product states.

\begin{figure*}[t]
\centering
\includegraphics[width=6in]{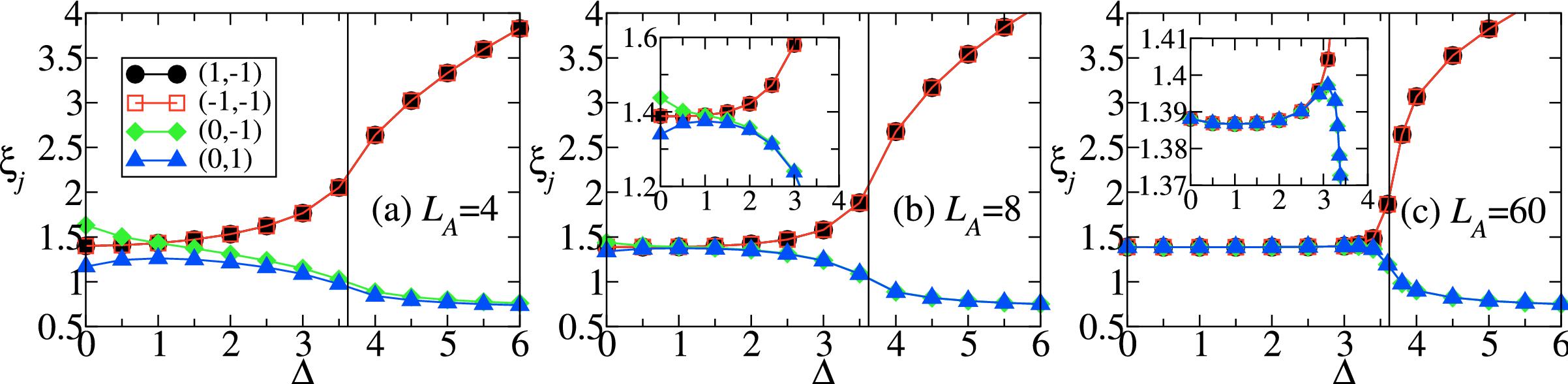} 
\caption{(a-c) The lowest four entanglement spectra, $\xi_j=-\ln\omega_j$, correspoding to Fig.~\ref{DLC-d0.3}(a-c). Each line is labelled with the quantum numbers $(S_A^z,p_A)$, where $S_A^z=0,\pm1$ and $p_A=\pm1$. See also the last paragrpah of the section ``Model Hamiltonian''. The insets of (b,c) are the zoom-in of (b,c). The vertical line denotes the critical point $\Delta_c\approx3.623$ obtained by the energy derivatives.}\label{ES-d0.3}
\end{figure*}

\begin{figure}[t]
\centering
\includegraphics[width=3in]{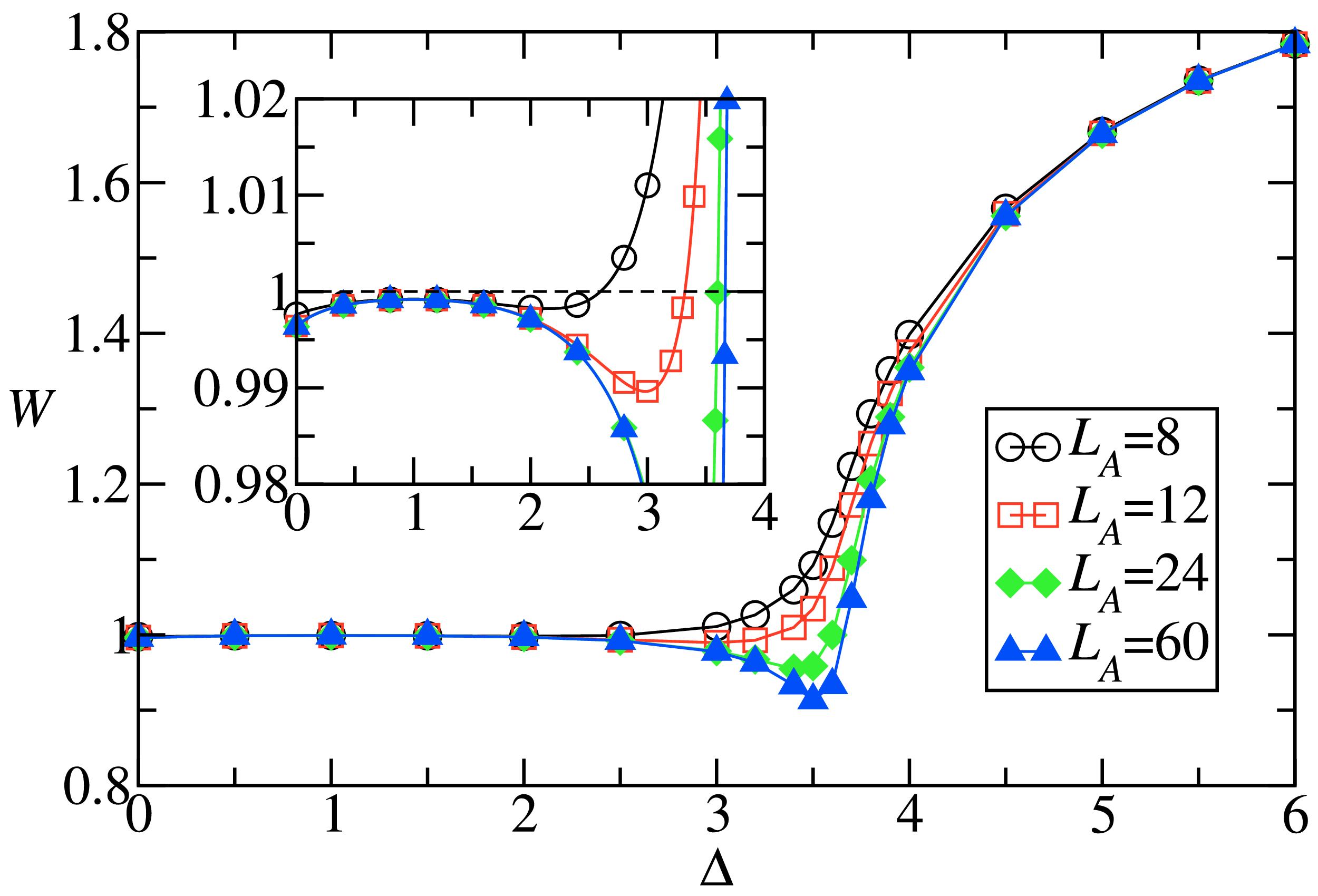} 
\caption{The quantity $W=4\sum_{j}\omega_j^2$ is plotted against $\Delta$ for $L_A=8,12,24,60$. The location of the local minima of W, corresponding to the local maxima of $S_2$, are defined as the pseudo-critical points $\Delta_2^*$ for the finite size scaling. The inset is the zoom-in plot.}\label{Fig4W-d0.3}
\end{figure}
 
In this paper, we focus on studying the Differential Local Convertibility (DLC), and study the edge states in the HD phase and the transitions from HD to the N\'eel phase, from CD to HD phase, as well as from CD to N\'eel phase and then to HD phase. Therefore, only the region $\Delta>0$ and $-1<\delta<1$ is present. See Fig.~\ref{fig:BAxxz}(b) for the four routes (the dashed lines) that will be swept along.

The DLC relies on the calculation of the correlation matrix of (\ref{BA-XXZ-H}).
By the Jordan-Wigner transformation, Eq.~(\ref{BA-XXZ-H}) can be mapping into spinless fermion chain. When $\Delta=0$, the model is exactly solvable~\cite{SSH,SQShen-TI-book,CFM-Peschel,free-fermion-CFM} for both finite $N$ and in the thermodynamic limit $N\to\infty$, see the section ``Methods'' for details. When $\Delta\neq 0$, we solve it numerically.
We use the Density Matrix Renormalization Group (DMRG) method~\cite{White1992} with the recently developed parity scheme (pDMRG).~\cite{Tzeng2012} 
%
%
%
The pDMRG is designed for using parity quantum numbers as well as freely selecting boundary conditions. Of course,  the effects of the boundaries vanish in thermodynamic limit. Here we comment, however,  that, besides  the general implications for a faster convergence to the thermodynamic limit and benefit for finite size scaling, our study based on DMRG with periodic boundary condition allows to examine the effects of the edge states arising from the bipartition of the system without the interference of the boundary effects.
%
%
%

Unlike the quantum Monte Carlo methods which may only simulate the R\'enyi entropy $S_\alpha$ with integer $\alpha\geq 2$,~\cite{QMC-1,QMC-2,QMC-3} the DMRG allows to compute $S_\alpha$ with a wide range of $\alpha$. As we will present in this work, $\alpha$ is taken from $10^{-2}$ to $10^3$. The pDMRG~\cite{Tzeng2012} enables us to label the eigenstates of $\rho_A$ by the quantum numbers $S_A^z$ and $p_A$, where $S_A^z=\frac{1}{2}\sum_{n\in A}\sigma_n^z$ is the $z$-component of total spins and $p_A$ is the parity (inversion) of the subsystem A. Thus the eigenvalues and eigenstates of $\rho_A$ can be identified by ($S_A^z, p_A$), which helps to better characterize the topological system.

\section*{Results}\label{Sec-DLC}
In this section, we shall present the results of DLC within each of the phase diagram Fig. \ref{fig:BAxxz}. In the first instance, we study the response of the ES to the sweep.
DLC is concerned with the convertibility between the ground state $\ket{\psi(g)}$ and $\ket{\psi(g+\epsilon)}$ of Eq.~(\ref{BA-XXZ-H}). The DLC can be related to the  derivative of the R\'enyi entropies: $\frac{\partial S_{\alpha}}{\partial g}$. If the latter is  non-negative, then $\ket{\psi(g)}$ can be converted to $\ket{\psi(g+\epsilon)}$, while the conversion changes direction otherwise. DLC breaks down if $\frac{\partial S_{\alpha}}{\partial g}\leq 0$ changes sign with $\alpha$. 

We also considered the majorization between $\ket{\psi(g)}$ and $\ket{\psi(g+\epsilon)}$.  In this way, we can study whether the catalyst is needed in the convertibility. The majorization between $\ket{\psi(g)}$ and $\ket{\psi(g+\epsilon)}$ is defined as 
\be
M(k)=\frac{\partial}{\partial g}\sum_{j=0}^k\omega_j,
\ee
where $\omega_j$'s are the eigenvalues of the reduced state of the subsystem. 
The local conversion between $\ket{\psi(g)}$ and $\ket{\psi(g+\epsilon)}$ without the aid of a catalyst is possible if the sign of $M(k)$ is uniform up to $0$ (see the section ``Introduction''). 

\subsection*{Sweeps along $\Delta$}
Here, we compute the DLC along the vertical sweep in Fig.~\ref{fig:BAxxz}. 

Fig.~\ref{DLC-d0.3}(a), (b), (c) show the sign of $\frac{\partial S_{\alpha}}{\partial\Delta}$ for $0\leq\Delta\leq 6$, $10^{-2}\leq\alpha\leq 10^{3}$, $\delta=0.3$ and $N=120$. The subsystem size $L_{A}=4,8,60$. Fig.~\ref{DLC-d0.3}(d), (e), (f) show the corresponding plots of majorization. We observe that, for a generic partition, the ground state in the HD cannot be converted. Nevertheless, we observe that the convertibility tends to be restored by increasing the partitions size. This phenomenon arises since the resources encoded in larger systems increase; therefore the convertibility is enhanced. This argument applies to one of the two subsystems whose size does not exceed $N/2$, as the effect of LOCC is more restricted by this subsystem rather than the other one (similar to the Schmidt decomposition of bipartite states). 
In view of this property, we fix the total size of the Hamiltonian and vary the size of subystem A. 

It can be seen in Fig.~\ref{DLC-d0.3}(a), (b), (c) that for $L_{A}=4,8$, DLC breaks down when $0\leq\Delta\lesssim3.6$, while it is positive when $\Delta\gtrsim3.6$. We notice that the critical point $\Delta_{c}\approx3.623$ separates the topological phase ($0\leq\Delta<\Delta_c$) from the non-topological N\'eel phase ($\Delta>\Delta_c$). It is also found that the inconvertible region in Fig.~\ref{DLC-d0.3}(c) shrinks when $L_{A}=N/2$ increases. We will examine it numerically in the section ``Methods'', where it is shown that the inconvertible region disappears in the thermodynamic limit.

Combining Figs.~\ref{DLC-d0.3}(c) with ~\ref{DLC-d0.3}(f), it can be seen that the {\it local conversion needs a catalyst in most part of the topological phase}, except around $\Delta=0$ and $\Delta=3$. In comparison, the catalyst is not necessary in the N\'eel phase. 
The local conversion changes direction at the SU(2) symmetry point $\Delta=1$ where the majorization shows a mirror-like symmetry. Moreover, it is conceivable that in the thermodynamic limit the conversion direction changes at the critical point $\Delta_{c}\approx3.623$. This is because the local conversion cannot increase the entanglement which diverges at the critical point (quantified by the von Neumann entropy, see Fig.~\ref{DLC-d0.3}(c) with $\alpha=1$).

\begin{figure*}[t]
\centering
\includegraphics[width=6in]{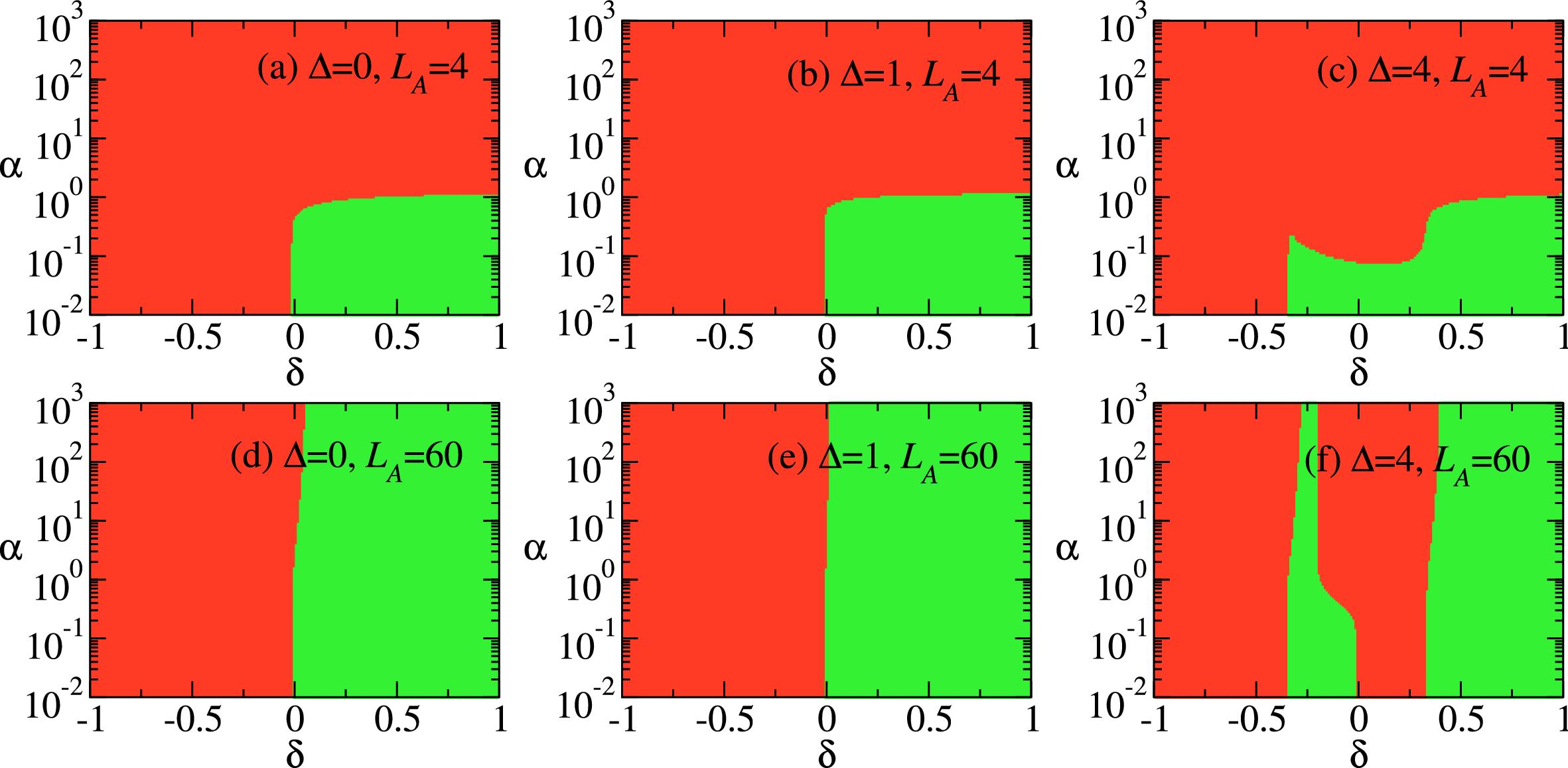}
\caption{\textbf{The sign of $\frac{\partial S_{\alpha}}{\partial\delta}$ plotted on the $\alpha$-$\delta$ plane.} Where $S_{\alpha}$ is the R\'enyi entropy of the reduced state of subsystem A. (a-c) The subsystem size $L_A=4$, $\Delta=0,1,4$. (d-f) $L_A=60$, $\Delta=0,1,4$. The total size is $N=120$ ($N\to\infty$ for $\Delta=0$). In the DMRG calculation, $m=300$ states are kept for $\Delta=4$, with the truncation error below $10^{-11}$ near the critical points, and below $10^{-14}$ away from the critical points.}\label{DLC-free}
\end{figure*}

In the following, we shall argue that the behavior of DLC can be understood in terms of edge states formation. Fig.~\ref{ES-d0.3}(a), (b), (c) show the lowest four eigenvalues of the entanglement Hamiltonian $H_E$, $e^{-H_E}=\rho_A$, for the three cases in Fig.~\ref{DLC-d0.3}. Let us consider $L_{A}=60$ first. It can be seen that the four eigenvalues of the reduced density matrix are almost degenerate in the HD phase. See also the zoom-in plot in Fig.~\ref{ES-d0.3}(c). Moreover, it is numerically found that the whole ES is at least four-fold degenerate. This suggests that the reduced state of subsystem A is approximately
\be\label{two-edge-states}
\rho_{A}\approx\begin{bmatrix}
\frac{1}{2}&0\\
0&\frac{1}{2}
\end{bmatrix}\otimes\rho_{0}\otimes\begin{bmatrix}
\frac{1}{2}&0\\
0&\frac{1}{2}
\end{bmatrix},
\ee
where $\rho_0$ is a $2^{L_A-2}\times 2^{L_A-2}$ matrix whose eigenvalues are equal to four times the eigenvalues of $\rho_A$. The two identical $2\times 2$ matrices in Eq.~(\ref{two-edge-states}) can be identified as the two edge states which induce the four-fold degeneracy of ES. This identification, manifesting the edge-ES correspondence,~\cite{edge-ES-correspondence-1,edge-ES-correspondence-2,edge-ES-correspondence-3} is supported by the two limiting cases below. (i) When $\delta=1$, the model is simplified as the sum of disconnected two-body \textit{XXZ} Hamiltonians. The ground state is a tensor product of spin singlet states.
If the subsystem A is chosen in such a way that both of its boundaries cut the singlet states, its reduced state is exactly the form in Eq.~(\ref{two-edge-states}).
The state $\rho_{0}$ is a pure state composed of singlet states of the bulk, while the other two identical matrices originate from the partial trace of the singlet states at the boundaries. (ii) When $\Delta=0$, Eq.~(\ref{BA-XXZ-H}) is equivalent to the Hamiltonian of non-interacting fermions, through the Jordan-Wigner transformation. See the ``Methods'' for details. In this case, the reduced state of the subsystem can be expressed as $\rho_{A}=e^{-\sum_{k}\varepsilon_{k}\tilde{c}_{k}^{\dagger}\tilde{c}_{k}}$ up to normalization. Here $\tilde{c}_{k}^{\dagger}$, $\tilde{c}_{k}$ are the creation, annihilation operators of the eigenmodes confined in the subsystem A with energy $\varepsilon_{k}$. When the subsystem is sufficiently long and its two boundaries cut the stronger bonds (with coupling strength $1+\delta$, $\delta>0$), there will be two eigenmodes which are localized about the boundaries with negligible energy, say, $\varepsilon_{1}\approx\varepsilon_2\approx 0$. As a result, the reduced state $\rho_A$ can be written in the form of Eq.~(\ref{two-edge-states}). For general cases ($\Delta\neq 0$ and $\delta\neq 1$), the ground state and the reduced state accordingly are complex many-body states. The four-fold ES degeneracy can be used to detect the edge states.~\cite{ES-topo-1}

\begin{figure*}[t]
\centering
\includegraphics[width=6in]{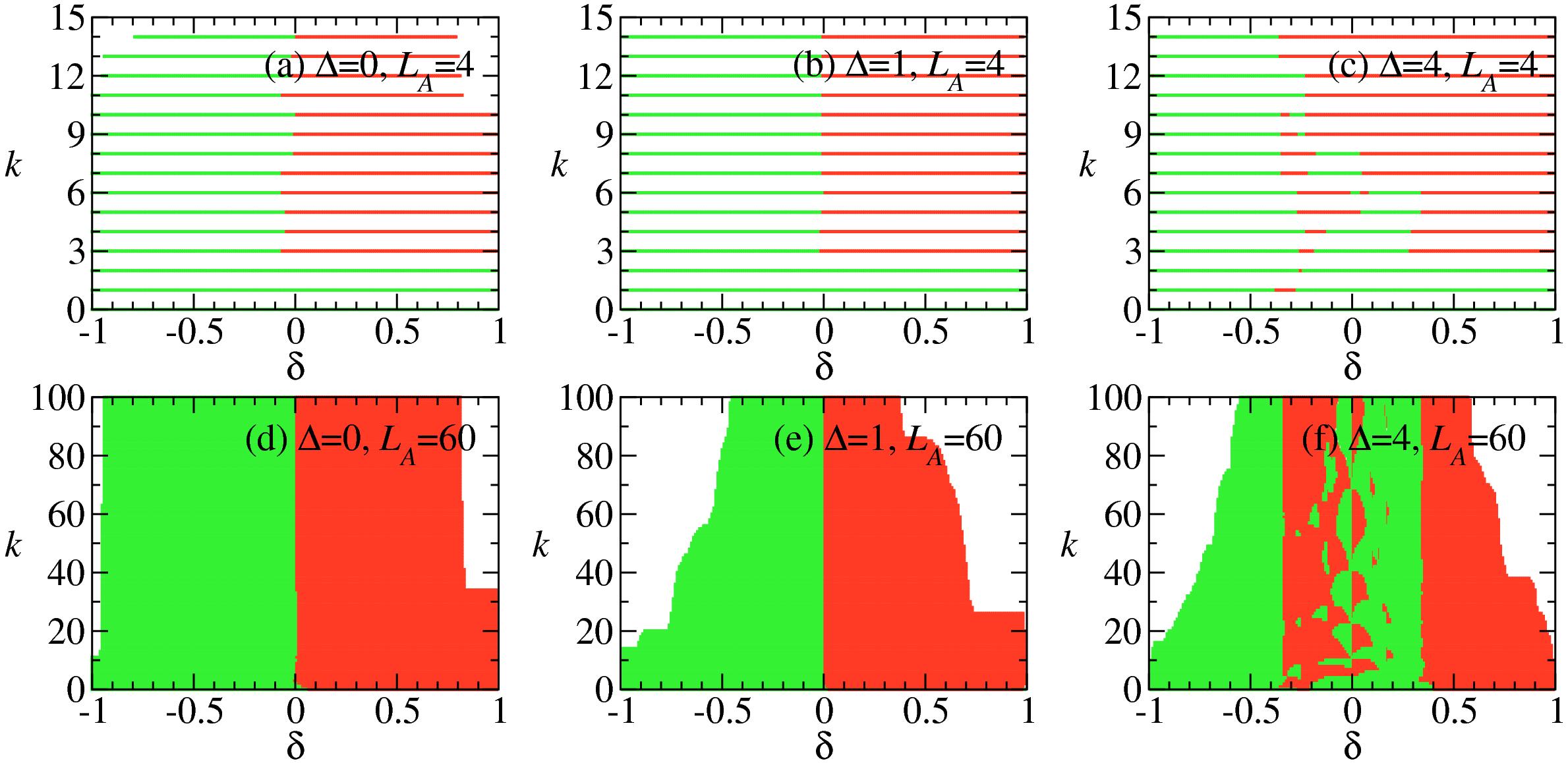}
\caption{\textbf{The majorization on the $k$-$\delta$ plane.} (a-f) are the six plots of majorization corresponding to the six plots in Fig.~\ref{DLC-free} respectively. The numerical error in the calculation of the eigenvalues of the reduced density matrix is below $10^{-13}$. The points with $|\sum_{j=0}^{k}\omega_j(\delta+\epsilon)-\omega_j(\delta)|<10^{-13}$ are not shown, where $\epsilon=5\times10^{-3}$. These points are mainly located at $|\delta|$ close to $1$.}\label{DLC-free-major}
\end{figure*}

In Fig.~\ref{Fig4W-d0.3}, the quantity $W=4\sum_j\omega_j^2$ is plotted vs $\Delta$, where $N=120$, $L_A=8,12,24,60$ and $\omega_j$'s are the eigenvalues of the reduced state of the subsystem A. The quantity $W$ with a large subsystem size basically shows the purity of the bulk $\rho_{0}$: $\mathrm{tr}(\rho_0^2)\approx 4\mathrm{tr}(\rho_A^2)=4\sum_j\omega_j^2$. It reflects the correlation between the bulk sites and the two edges, as interpreted below. 
Fig.~\ref{Fig4W-d0.3} shows that the purity of the bulk state increases with $\Delta$ to reach a maximum at $\Delta=1$, and then it starts to decrease. This behavior indicates that the bulk sites become less correlated with the two edges when $\Delta$ increases to approach $1$, while they do the opposite when $\Delta>1$. This point can be understood by considering the special case $\delta=1$ where the bulk state is a pure state composed of singlet states so that the bulk sites are completely uncorrelated with the edges. Therefore, the lowest four eigenvalues of $H_E$ have a local minimum at $\Delta=1$, where they are closest to $-\ln0.25\approx1.386$. See the inset of Fig.~\ref{ES-d0.3}(c). The bifurcation occurs when $\Delta\gtrsim3.1$, showing that the four-fold ES degeneracy is partially lifted. As a consequence, Eq.~(\ref{two-edge-states}) is no longer applicable. It's conceivable that the edge states start recombining, resulting in the splitting of the lowest four entanglement spectra. In particular, $\xi_0\equiv-\ln\omega_0$ will decrease with $\Delta$.

The above analysis is consistent with DLC in Fig.~\ref{DLC-d0.3}(c). Since $\lim\limits_{\alpha\to\infty}S_{\alpha}(\rho_{A})=-\ln\omega_0$, the sign of $\frac{\partial S_{\alpha}}{\partial\Delta}$ for large $\alpha$ in Fig.~\ref{DLC-d0.3} can be derived from Fig.~\ref{ES-d0.3}(f). In addition, when $\alpha\to 1$, the R\'enyi entropy reduces to the von Neumann entropy which has been shown to diverge at critical points.~\cite{vn-entropy-for-phase-1,vn-entropy-for-phase-2,vn-entropy-for-phase-3} This result was verified in the spin-1/2 XXZ chain with bond alternation.~\cite{Tian2013}

Next, we consider $L_A=4,8$. Comparing Fig.~\ref{DLC-d0.3}(a), (b) with (c), we find that the inconvertible region in Fig.~\ref{DLC-d0.3}(c) increases when the subsystem size decreases. Namely, the ground state changes from convertible to non-convertible at $\Delta$ that becomes smaller when $L_A$ decreases. This indicates that the edge states start recombining at smaller $\Delta$ when $L_A$ decreases. 
Note that $\Delta=1$ no longer separates the positive and negative DLC regions: all the regions in HD phase are locally inconvertible due to the recombination of edge states for small $L_A$. 

In the final part of the present subsection, we shall give a physical interpretation for Fig.~\ref{ES-d0.3}(a), (b). Roughly speaking, the term $\sum_{n}\Delta\sigma_{n}^{z}\sigma_{n+1}^{z}$ tends to anti-parallel the $z$ components of neighboring spins, which has an effect that the $z$ components of spins of the two edges of subsystem A are also anti-parallel. Thus, the probability $\omega_j$ of the eigenstates with quantum number $(S_A^z,p_A)=(0,-1)$ of $\rho_A$ increases with $\Delta$. These eigenstates are the anti-parallel component of the triplet states with $p_{A}=-1$. Based on this picture, the change of ES ($\xi_{j}=-\ln\omega_{j}$) when $\Delta$ is varied, as shown in Fig.~\ref{ES-d0.3}(a), (b), can be understood. Both the Schmidt states with quantum number $(0,-1)$ and those with $(0,1)$ have anti-parallel $z$ components of spins for the two edges. The reason why the term $\sum_{n}\Delta\sigma_{n}^{z}\sigma_{n+1}^{z}$ prefers the former to the latter when $0\leq\Delta\leq1$ is related to the ES level crossing at the SU(2) symmetry point $\Delta=1$. We notice that due to the level crossing, the sign of the derivative of R\'enyi entropy flips at $\Delta=1$ for all $\alpha$. This means that the ES of the singlet state with quantum number $(0,1)$ has an extreme point at $\Delta=1$ and around it a mirror-like symmetry is present.~\cite{DLC-XXZ-non-universal}

The reduced state of the special case $\Delta=0$ is shown in Eq.~(\ref{appendix-free-fermion-edge}). It can be seen in Fig.~\ref{ES-d0.3}(a), (b) that the ES splitting becomes smaller when $\Delta$ increases from 0 to 1, and then it becomes larger when $\Delta$ increases further. These results indicate that there is a minimum recombination of the edge states at $\Delta=1$. Beyond the critical point, the edge states disappear and the system goes to the N\'eel phase. In this phase the ground state is convertible.

\begin{figure*}[t]
\centering
\includegraphics[width=6in]{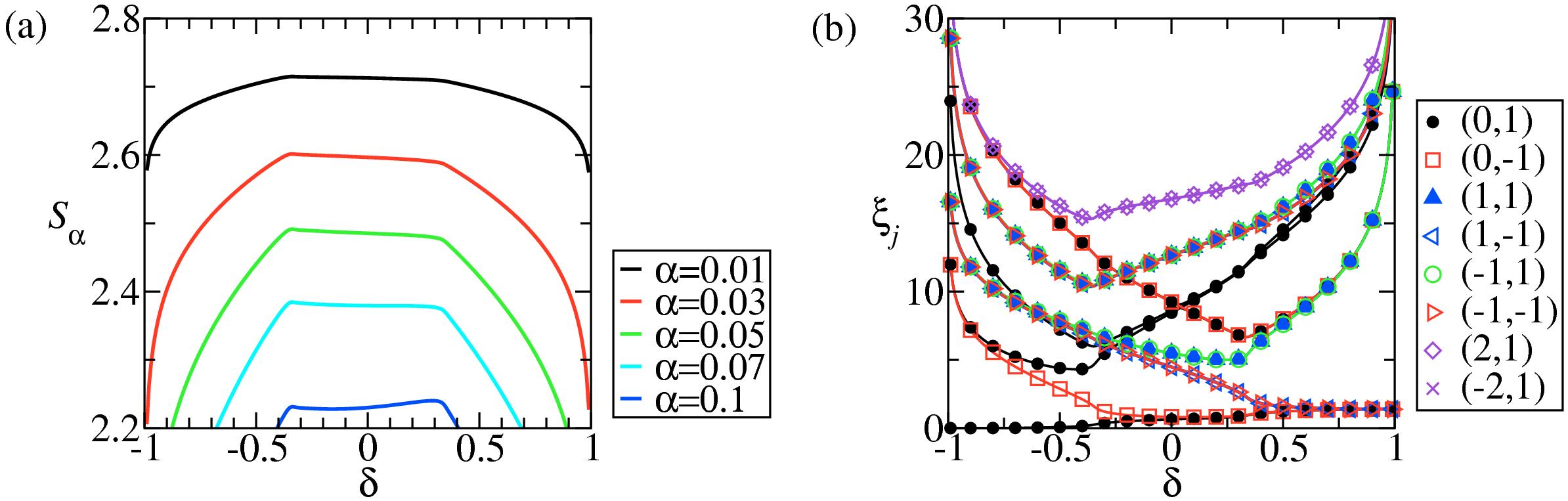} 
\caption{\textbf{The small-$\alpha$ R\'enyi entropies and the entanglement spectra for $\Delta=4$, $L_A=4$ and $N=120$.} (a) The small-$\alpha$ R\'enyi entropies ($\alpha=0.01,0.03,0.05,0.07,0.1$) and (b) the entanglement spectrum, $\xi_j=-\ln\omega_j$, corresponding to Fig.~\ref{DLC-free}(c). The number pairs in the legend denote the values of the quantum numbers $(S_A^z,p_A)$.}\label{free-Renyi-ES}
\end{figure*}

\subsection*{Sweeps along $\delta$}

Fig.~\ref{DLC-free} shows the sign of $\frac{\partial S_{\alpha}}{\partial\delta}$ for $-1\leq\delta\leq 1$, $10^{-2}\leq\alpha\leq 10^{3}$, $\Delta=0,1,4$ and $L_A=4,60$. The case $\Delta=0$ is calculated by using the correlation matrix formalism in the section ``Methods'' with $N\to\infty$. For $\Delta=1,4$, DMRG is used and $N=120$. The topological regime is $0<\delta\leq1$ for $\Delta=0,1$, and $0.34\lesssim\delta\leq1$ for $\Delta=4$.

Fig.~\ref{DLC-free} shows that DLC cannot be achieved within  the topological phase, at small sizes $L_A$. These results can be interpreted by the recombination of edge states (see the previous subsection). The inconvertible region in the topological phase shrinks when $L_A$ increases, and it will disappear in the thermodynamic limit (this will be discussed in the subsection ``Methods: Phase diagram''). This phenomenon was also discussed in the previous subsection. In contrast with the sweep along $\Delta$, however, the local conversion in {\it part of the convertible topological phase does not need the catalyst}. See Fig.~\ref{DLC-free-major} for the majorization. This indicates  that in the large $L_A$ limit, the phases cannot be told apart   neither by looking at the necessity of the catalyst.

In contrast with the previous sweep, the DLC can be violated within the  N\'eel phase, Fig.~\ref{DLC-free}(c), and Fig.~\ref{DLC-free}(f). 
As detailed below, such a phenomenon arises because of the two consecutive continuous quantum phase transitions bounding the N\'eel phase.
The sweep along $\delta$ goes through three phases: CD, N\'eel and HD. Since the R\'enyi entropies diverge at the two critical points $\delta\approx\pm0.34$ for infinite $L_A$, there must be two maxima respectively around the two points for large $L_A$. Apparently, a minimum between the two maxima is present for every $\alpha$, as can be seen in Fig.~\ref{DLC-free}(f). The value of $\delta$ corresponding to this minimum in general varies with $\alpha$ (unless some additional symmetry is present in the subsystem A, like the SU(2) at $\Delta=1$ in Fig.~\ref{DLC-d0.3}(c), but it does not seem to exist here). As a result, that specific   N\'eel phase may result unconvertible. For small $L_A$, the value of $\alpha$ corresponding to a negative derivative of the R\'enyi entropy is small: $10^{-2}\lesssim\alpha\leq10^{-1}$. See Fig.~\ref{DLC-free}(c). This is understood as the residual influence of the above three-phase mechanism on the R\'enyi entropies. Fig.~\ref{free-Renyi-ES}(a) shows the R\'enyi entropy for various values of small $\alpha$. It can be seen that in the N\'eel phase, the R\'enyi entropy decreases very slowly for $10^{-2}\lesssim\alpha<10^{-1}$. In contrast, it decreases rapidly in the HD phase. 

The behavior of the small-$\alpha$ R\'enyi entropies can be understood by inspecting the entanglement spectrum, as shown in Fig.~\ref{free-Renyi-ES}(b). It can be seen that in the N\'eel phase, the eigenvalues of the entanglement Hamiltonian, $\xi_j$'s, are smaller than $18$. The corresponding eigenvalues of $\rho_A$ are larger than $e^{-18}\approx 1.5\times 10^{-8}$. Thus, the rank of $\rho_A$ is $16$ which is unchanged with $\delta$. Also, the large $\xi_j$'s increase slowly with $\delta$, as compared with their change in the HD phase. Some of them even decreases. Since the small-$\alpha$ R\'enyi entropies are susceptible to the large $\xi_j$'s, their slow decrease with $\delta$ is understood. In the HD phase, the large $\xi_j$'s increases rapidly with $\delta$, resulting in a rapid decrease of the effective rank of $\rho_A$. The rapid decrease is a consequence of the formation of the edge states when $\delta$ approaches $1$: the small eigenvalues of $\rho_A$ disappears and only the four largest eigenvalues dominates. Therefore, the small-$\alpha$ R\'enyi entropies, representing the effective rank of $\rho_A$, also decrease rapidly with $\delta$.

\section*{Discussion}\label{Sec-conclusion}
We have investigated the entanglement convertibility in the one-dimensional spin-1/2 \textit{XXZ} model with bond alternation. The phase diagram is parametrized by the Ising-type anisotropy $\Delta$ and the bond alternation $\delta$ (see Fig. \ref{fig:BAxxz}(b)). We sweep both in $\Delta$ (fixed $\delta$) and in $\delta$ (fixed $\Delta$). The method we exploit for calculating the R\'enyi entropies is the pDMRG~\cite{Tzeng2012} and the correlation function matrix formalism~\cite{CFM-Peschel,free-fermion-CFM}, applied to systems with periodic boundary condition. Such calculations are carried out by a bipartition of the system A$\mid$B with blocks of length $L_A$, $L_B$, and tracing out B. The finite-size scaling of the maximum of R\'enyi entropies $S_\alpha$ with $\alpha\to\infty$ and $\alpha=2$ are used to locate the critical point. It is compared with the finite-size scaling of second derivatives of ground state energy density and they agree well. The precise ground-state phase diagram is determined.

Our results confirm that the response of the entanglement spectrum is markedly different in topological and non-topological phases: The effective rank of the reduced density matrix changes much faster in the topological phase than in the N\'eel phase. As detailed below, such a phenomenon is responsible for the violation of DLC within the topological Haldane dimer phase at small $L_A$.
%
%
%
Our study demonstrates how the response of the entanglement spectrum to the Hamiltonian parameters implies a non trivial property for the convertibility of the ground state with Haldane order.  Such a phase is also found to be characterized by a ''large susceptibility'' of the effective rank of the reduced density matrix of the sub-system.
%
%
%

\textbf{\textit{Differential Local Convertibility.}}
Our results confirm that the DLC depends on the ratio between the spin-spin correlation length and the size of the partition. In agreement with the Ref.~\onlinecite{DLC-XXZ-non-universal}, our results obtained in the case of large subsystems indicate that 
the direction of local conversion is reversed at the SU(2) symmetry point. 
In agreement with the Ref.~\onlinecite{DLC-Cluster-XXZD}, for a small subsystem size,  the state results unconvertible within the SPTO Haldane dimer phase,  for both sweeps in  $\Delta$ and $\delta$. Such a result arises as a recombination of edge states as in Ref.~\onlinecite{DLC-Ising}. The ground state results DLC within the classical Dimer phase. In contrast with the case analyzed in previous studies, however, the DLC changes within the N\'eel phase.  This phenomenon is a specific effect of the presence of the two continuous quantum phase transitions bounding the N\'eel phase: Since the R\'enyi entropies diverge at the two critical points $\delta_c$ for infinite $L_A$, there must be two maxima respectively around the two points for large $L_A$. Apparently, a minimum between the two maxima is present for every $\alpha$, as can be seen in Fig.~\ref{DLC-free}(f). The value of $\delta$ corresponding to this minimum in general varies with $\alpha$ (unless some additional symmetry is present in the subsystem A, like the SU(2) at $\Delta=1$ in Fig.~\ref{DLC-d0.3}(c), but it does not seem to exist here). As a result, negative DLC emerges in the N\'eel phase. 

\textbf{\textit{Majorization.}}
We find the conversion fulfills the majorization relation in the non-topological phases.
In the sweep along $\delta$ in the topological phase, the conversion violates the majorization relation, even in the large $L_A$ limit. 
This result indicates that the catalyst is strictly necessary in the process.
In the sweep along  $\Delta$, however, the catalyst seems not needed in the part of the topological phase.

Overall, our study can provide a characterization of the topological order through the response of the entanglement spectrum, on a local scale. This may facilitate the experimental sought of topological order~\cite{nature2015,measure-Renyi-entropy-1,measure-Renyi-entropy-2}. On the more quantum information side, our results contribute to the questions on whether topological phases universally encode more computational power than non-topological phases. 


\section*{Methods}\small
\subsection*{Dimerized chain of non-interacting spinless fermions}\label{appendix-free-fermion}
When $\Delta=0$, the Hamiltonian in Eq.~(\ref{BA-XXZ-H}) can be mapped to the dimerized chain of non-interacting spinless fermions~\cite{SSH} through the Jordan-Wigner transformation, and the entanglement spectrum can be computed using the correlation function matrix (CFM) formalism.~\cite{CFM-Peschel,free-fermion-CFM}

The Jordan-Wigner transformation is
\ba\label{appendix-JW}
c_{j}=\sigma_{j}\prod_{k=1}^{j-1}(-\sigma_{k}^{z}),
\ea
where $\sigma_{j}=(\sigma_j^x-i\sigma_j^y)/2$. Substituting Eq.~(\ref{appendix-JW}) into Eq.~(\ref{BA-XXZ-H}), we have
\begin{align}\label{H-free-fermion}
H&=-2\sum_{n=1}^{N-1}[1+(-1)^n\delta](c_{n}c_{n+1}^{\dagger}+c_{n+1}c_{n}^{\dagger})\nonumber\\
&+2[1+(-1)^N\delta](c_{N}c_{1}^{\dagger}+c_{1}c_{N}^{\dagger})\prod_{j=1}^{N}(1-2c_{j}^{\dagger}c_{j}).
\end{align}

Assume $N$ is an even number and $M=N/2$. Since the number operator $\sum_{j}c_{j}^{\dagger}c_{j}$ commutes with the Hamiltonian, the eigenstates of the Hamiltonian can be solved in the subspace of conserved number of particles. In particular, it will be shown later that the ground state is non-degenerate and contains $M$ particles (\textit{i.e.} half-filled) when $\delta\neq0$. In this case, the operator $\prod_{j=1}^{N}(1-2c_{j}^{\dagger}c_{j})$ in Eq.~(\ref{H-free-fermion}) is equal to $(-1)^M$. Thus, we have a free fermion chain with periodic boundary conditions (PBC) when $M$ is odd, while the boundary condition is anti-periodic when $M$ is even.

Perform the Fourier transformation
\ba\label{appendix-Fourier}
c_{2j-1}=\frac{1}{\sqrt{M}}\sum_{k=1}^{M}a_{k}e^{i\frac{p_{k}\pi}{M}j},\\
c_{2j}=\frac{1}{\sqrt{M}}\sum_{k=1}^{M}b_{k}e^{i\frac{p_{k}\pi}{M}j},
\ea
where $p_{k}$ depends on $M$: when $M$ is odd, $p_{k}=2k$, otherwise $p_{k}=2k+1$. The choice of $p_{k}$ is to ensure that the Fourier transformation works for both PBC and anti-PBC fermion chains.
The Hamiltonian is transformed into
\ba
H=\sum_{k=1}^{M}\begin{bmatrix}
a_k^\dagger\hspace{3mm}b_k^\dagger
\end{bmatrix}[\boldsymbol{R}(k)\cdot\boldsymbol{\sigma}]\begin{bmatrix}
a_k\\b_k
\end{bmatrix},
\ea
where $\boldsymbol{R}(k)=(R_{x}(k),R_{y}(k),R_{z}(k))$, $R_{x}(k)=2(1-\delta)+2(1+\delta)\cos\frac{p_k\pi}{M}$, $R_{y}(k)=2(1+\delta)\sin\frac{p_k\pi}{M}$, $R_{z}(k)=0$, and $\boldsymbol{\sigma}=(\sigma_{x}\sigma_{y},\sigma_{z})$ is the vector of Pauli matrices. The length of $\boldsymbol{R}(k)$ is $R(k)=4\sqrt{\cos^{2}\frac{p_{k}\pi}{2M}+\delta^{2}\sin^{2}\frac{p_{k}\pi}{2M}}$. The Hamiltonian has two bands with energy $\pm R(k)$. When $\delta\neq0$, $R(k)$ is nonzero for all $k$ and the two bands are gapped. The ground state corresponds to the occupied negative-energy band ( half-filled) and it is non-degenerate. When $\delta=0$, the Hamiltonian is gapless, since the two bands touch when $M\to\infty$ ($R(k)\to 0$ for $p_k/M\to 1$).

The topological properties of the chain can be characterized by the Berry phase~\cite{SQShen-TI-book} $\gamma\equiv\int_{0}^{2\pi}dk\bra{\phi}i\partial_k\ket{\phi}$, where $\ket{\phi}$ is the eigenvector of $\boldsymbol{R}(k)\cdot\boldsymbol{\sigma}$ with the eigenvalue $-R(k)$. It can be shown that $\gamma=n_{w}\pi$ modulo $2\pi$, where $n_{w}\equiv\frac{1}{2\pi}\int_{0}^{2\pi}\frac{d\theta}{dk}dk=[1+\textrm{sign}(\delta)]/2$. Here $\theta$ is the polar angle: $\tan(\theta)=R_y(k)/R_x(k)$. The quantity $n_{w}$ is the winding number describing the total number of times that $\boldsymbol{R}(k)$ surrounds the origin of the $(R_x(k),R_y(k))$ parametric space when $k$ changes from $0$ to $2\pi$. A nonzero $n_w$ (\textit{i.e.} $\delta>0$) defines the topological phase, where the chain with open-boundary conditions supports two edge modes. For PBC, the reduced state of subsystem has two edge modes when two stronger bonds are cut off. This can be derived by the CFM below.

The CFM is defined as $C_{m,n}=\bra{\phi_0}\boldsymbol{c}_{m}\boldsymbol{c}_{n}^{\dagger}\ket{\phi_0}$, where $\ket{\phi_0}$ is the ground state and $\boldsymbol{c}_{m}=\begin{bmatrix}
c_{2m-1}\hspace{3mm}c_{2m}
\end{bmatrix}^T$. It can be verified that $C_{m,n}=\frac{1}{M}\sum_{k}e^{i\frac{p_{k}\pi}{M}(m-n)}G(k)$, where $G(k)$ is the CFM in momentum space:~\cite{free-fermion-CFM}
\ba
G(k)=\frac{1}{2}[I_{2\times2}+\hat{\boldsymbol{R}}(k)\cdot\boldsymbol{\sigma}],
\ea
where $\hat{\boldsymbol{R}}(k)=\boldsymbol{R}(k)/R(k)$ is a unit vector. The reduced state of subsystem A is $\rho_{A}=\frac{1}{Z}e^{-\sum_{l}\varepsilon_{l}\tilde{c}^{\dagger}_{l}\tilde{c}_{l}}$, where $Z=\textrm{tr}(e^{-\sum_{l}\varepsilon_{l}\tilde{c}^{\dagger}_{l}\tilde{c}_{l}})$ is the normalization constant, $\varepsilon_{l}=\ln\frac{q_l}{1-q_l}$, $\tilde{c}_{l}=\sum U_{lm}c_{m}$, and $U$ is the unitary matrix that diagonalizes the CFM with eigenvalues $q_{l}$ (\textit{i.e.} $U[C_{m,n}]U^{\dagger}$ is a diagonal matrix). The reduced state $\rho_A$ can be written compactly as~\cite{reduced-state-compact} $\rho_A=\textrm{det}(I-C)e^{\boldsymbol{c}^{\dagger}\ln[C(I-C)^{-1}]\boldsymbol{c}}$, where $I$ is the $L_A\times L_A$ identity matrix and $C\equiv[C_{m,n}]$.

When the subsystem is defined by cutting off two stronger bonds, there will be two zero-energy edge modes, say $q_{1}\approx q_{2}\approx\frac{1}{2}$. We have
\begin{align}\label{appendix-free-fermion-edge}
\rho_{A}&=\rho_{0}\otimes\frac{1}{Z_{1}}e^{-\varepsilon_{1}(\tilde{c}_{1}^{\dagger}\tilde{c}_{1}-\tilde{c}_{2}^{\dagger}\tilde{c}_{2})}\nonumber\\
&=\begin{bmatrix}
\frac{1}{2}-\lambda&0\\
0&\frac{1}{2}+\lambda
\end{bmatrix}\otimes\rho_{0}\otimes\begin{bmatrix}
\frac{1}{2}+\lambda&0\\
0&\frac{1}{2}-\lambda
\end{bmatrix},
\end{align}
where $\rho_{0}$ is the reduced state of all the eigen-modes excluding the edge modes, $Z_{1}=\textrm{tr}(e^{-\varepsilon_{1}(\tilde{c}_{1}^{\dagger}\tilde{c}_{1}-\tilde{c}_{2}^{\dagger}\tilde{c}_{2})})$,  $\varepsilon_{1}$ is positive and equals $\ln\frac{q_1}{1-q_1}\approx0$, $\lambda=q_1-\frac{1}{2}$, $\tilde{c}_{1}^{\dagger}=\frac{1}{\sqrt{2}}(\tilde{c}_{L}^{\dagger}-\tilde{c}_{R}^{\dagger})$, $\tilde{c}_{2}^{\dagger}=\frac{1}{\sqrt{2}}(\tilde{c}_{L}^{\dagger}+\tilde{c}_{R}^{\dagger})$. Here, $\tilde{c}_{L}^{\dagger},\tilde{c}_{R}^{\dagger}$ are the creation operators of the left and right edge modes. The wave function for the edge modes can be derived from the eigenvectors of $\rho_{A}$ (corresponding to the eigenvalues $q_{1,2}$). The form of $\tilde{c}_{1,2}^{\dagger}$ is consistent with the requirement that $\rho_{A}$ commutes with the inversion symmetry operator within subsystem A. Note that Eq.~(\ref{appendix-free-fermion-edge}) is a special case of Eq.~(\ref{two-edge-states}). Also, the two degenerate entanglement spectra at $\Delta=0$ in Fig.~\ref{ES-d0.3}(a) can be understood by inspecting the matrix elements of the two edge states in Eq.~(\ref{appendix-free-fermion-edge}).

The R\'enyi entropy of $\rho_A$ when $\Delta=0$ is simplified as
\ba
S_\alpha(\rho_A)=\frac{1}{1-\alpha}\sum_{j=1}^{L_A}\ln[q_j^\alpha+(1-q_j)^\alpha].
\ea

%

\subsection*{Phase diagram}\label{sec:phase}

\begin{figure}[t]
\centering
\includegraphics[width=2.8in]{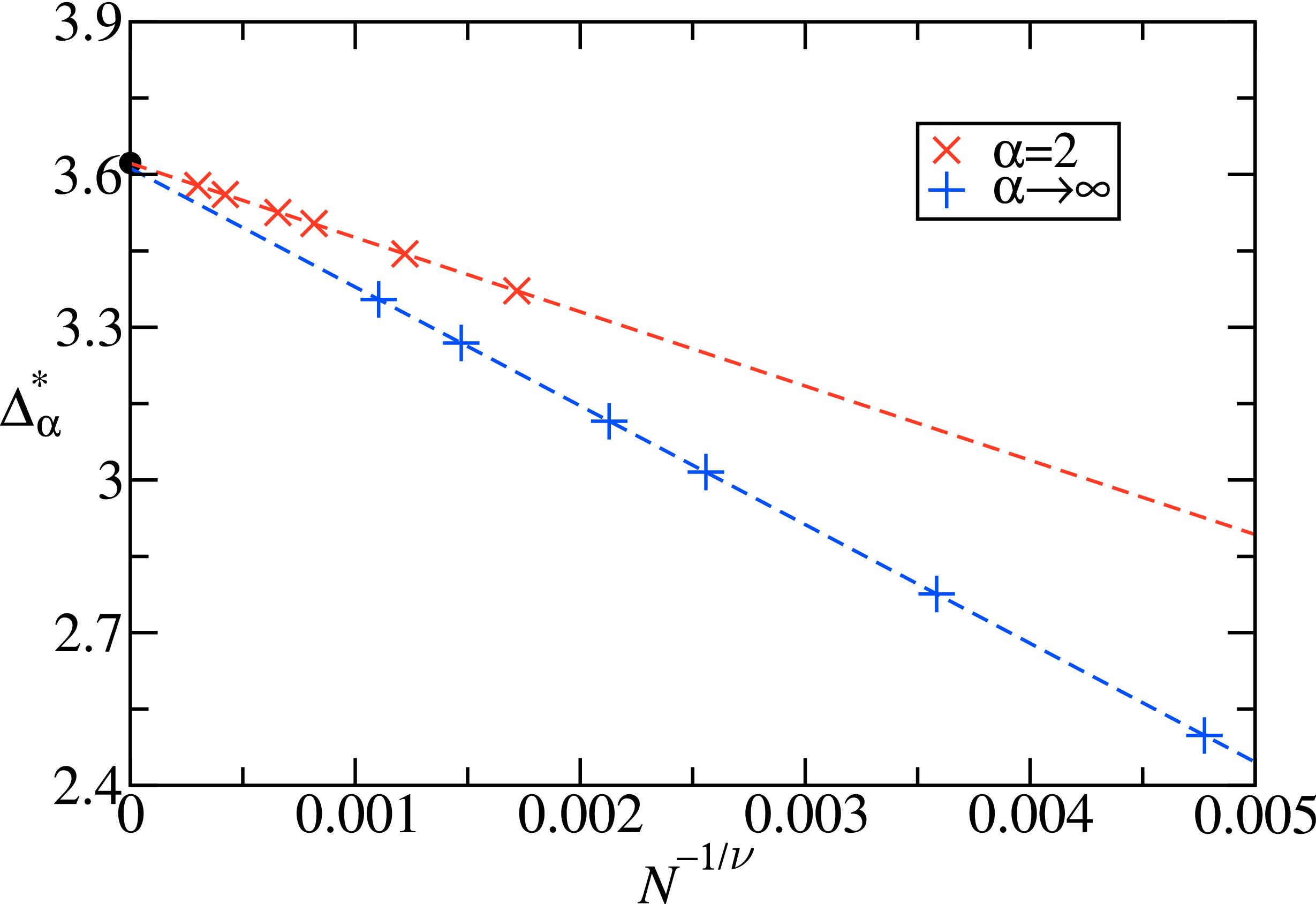}
\caption{Extrapolation of critical points from the R\'enyi entropies with $\alpha=2$ and $\alpha\to\infty$ for the sweeps along $\Delta$ while fixing $\delta=0.3$. The $S_2$ is related to the quantum purity which can be measured directly in the experiment,~\cite{nature2015} and $S_\infty =\xi_0$ is the lowest entanglement spectrum. The critical point $\Delta_c\approx3.622$ and $3.612$ for $\alpha=2$ and $\alpha\to\infty$, with the exponents $\nu\approx0.653$ and $0.778$, respectively. The black point denotes the critical value of $\Delta_c\approx3.623$ obtained from the 2nd derivatives of ground state energy.}\label{fig:scaling}
\end{figure}

In this subsection, we describe the methodology for getting precise phase diagram of the model Hamiltonian. We have shown in the section ``Results'' that the topological states are not convertible for a small subsystem $L_A=4$. However, since the non-topological phase there may also present negative convertibility, one can not in general detect the topological transitions by examining the boundary of convertibility. We therefore look for other quantities in the case of half-half bipartition, $L_A=N/2$, and performing extrapolation to the thermodynamic limit $N\to\infty$. 
%
%
%
The known literature demonstrates that R\'enyi entropies are logarithmically divergent in Luttinger liquids~\cite{von-logL,Renyi-entropy-diverge}. Here, we detect the critical points via the analysis of the specific Luttinger liquids encoded in the quantum critical regimes of our system~\cite{Renyi-entropy-diverge} through the finite size scaling theory of the Renyi entropies $S_\alpha$.
Similar scaling technique for the von Neumann entropy has been applied to finding quantum critical points in spin-1 system.~\cite{Tzeng2008a,Tzeng2008b}
We concentrate on two special cases of $S_\alpha$: $\alpha\to\infty$ and $\alpha=2$, as they are closely related to DLC in our discussion. Other values of $\alpha$ can also be considered, in principle.
%
%
%

The sign-changed point $\Delta^*_\infty$ of DLC for infinite $\alpha$, as pointed out in the Fig.~\ref{DLC-d0.3}(c). This point indicates the beginning of the recombination of edge states. We refer to this point as the pseudo-critical point for a finite $N$. When $N\to\infty$, it converges to the infinite system's critical point. We numerically determine the pseudo-critical point $\Delta_\infty^*$ such that $\partial_\Delta S_{\infty}=\partial_\Delta\xi_0=0$, where $\xi_0$ is the lowest entanglement spectrum.
%
%
%
We note that the Schmidt gap, $G=\omega_0-\omega_1$, or equivalently, the entanglement gap, $\Delta\xi=\xi_1-\xi_0$, is used for characterizing quantum phase transitions recently,~\cite{SchmidtGap1,SchmidtGap2} however we only consider $S_\infty=\xi_0$
%
and provide a new attempt to locate quantum critical points from the entanglement spectrum.
%
%
When the system size $N$ increases, the shift of this point represents the shrink of the region of negative convertibility. As shown in Fig.~\ref{fig:scaling}, $\Delta_\infty^*$ approaches the critical point $\Delta_c\approx3.612$ in the thermodynamics limit $N\to\infty$. In fact, the R\'enyi entropies exhibit logarithmic divergence: $S_\alpha\propto\ln L_A$  in an infinite gapless one-dimensional model.~\cite{Renyi-entropy-diverge} Thus, the extreme point of R\'enyi entropies must converge to the same critical point when $N\to\infty$. Now we have more confidence to say that, for fixed $\delta$ and varying the anisotropy parameter $\Delta$ with half-half bipartition, both the HD phase and N\'eel phase have positive convertibilities in the thermodynamic limit.

We now consider the R\'enyi entropy with $\alpha=2$. There are three advantages. (i) It reflects the purity of $\rho_A$ as $S_2=-\ln\textrm{tr}(\rho_A^2)$. It is also related to the quantity $W$ in Fig.~\ref{Fig4W-d0.3} and thus the purity of the bulk state $\rho_0$ when Eq.~(\ref{two-edge-states}) is valid: $S_2=-\ln\frac{W}{4}\approx-\ln\frac{\textrm{tr}(\rho_0^2)}{4}$. (ii) $S_2$ can be measured directly in experiments without reconstructing the eigenvalues of $\rho_A$.~\cite{measure-Renyi-entropy-1,measure-Renyi-entropy-2,nature2015} (iii) In the quantum Monte Carlo methods, for $\alpha=1$, the von Neumann entropy $S_{\rm{v}}$ is difficult to simulate. R\'enyi entropies with integer $\alpha\geq2$, especially $S_2$ is easer to be simulated.~\cite{QMC-1,QMC-2,QMC-3} We numerically determine the pseudo-critical point $\Delta_2^*$ by locating the maximum of $S_2$, which is also the minimum of $W$ in Fig.~\ref{Fig4W-d0.3}. As shown in Fig.~\ref{fig:scaling}, for $\delta=0.3$ the critical point $\Delta_c\approx3.622$ is obtained precisely.

However, when very small value of $\delta\approx 0^+$ is fixed, the HD phase are close to the critical Luttinger liquid for the region $\Delta\lesssim 1$, and the correlation lengths are large. The typical DLC, as in Fig.~\ref{DLC-d0.3}(c), should only appear when the subsystem size is much larger than the correlation length. This makes numerical difficulty for finding the proper pseudo-critical point $\Delta_\alpha^*$.
The precise ground-state phase diagram for the bond alternating XXZ model, shown in the Fig.~\ref{fig:BAxxz}(b), is determined by the above two methods and the scaling of the second derivatives of ground state energy density which is discussed below.

\begin{figure}[t]
\centering
\includegraphics[width=3.3in]{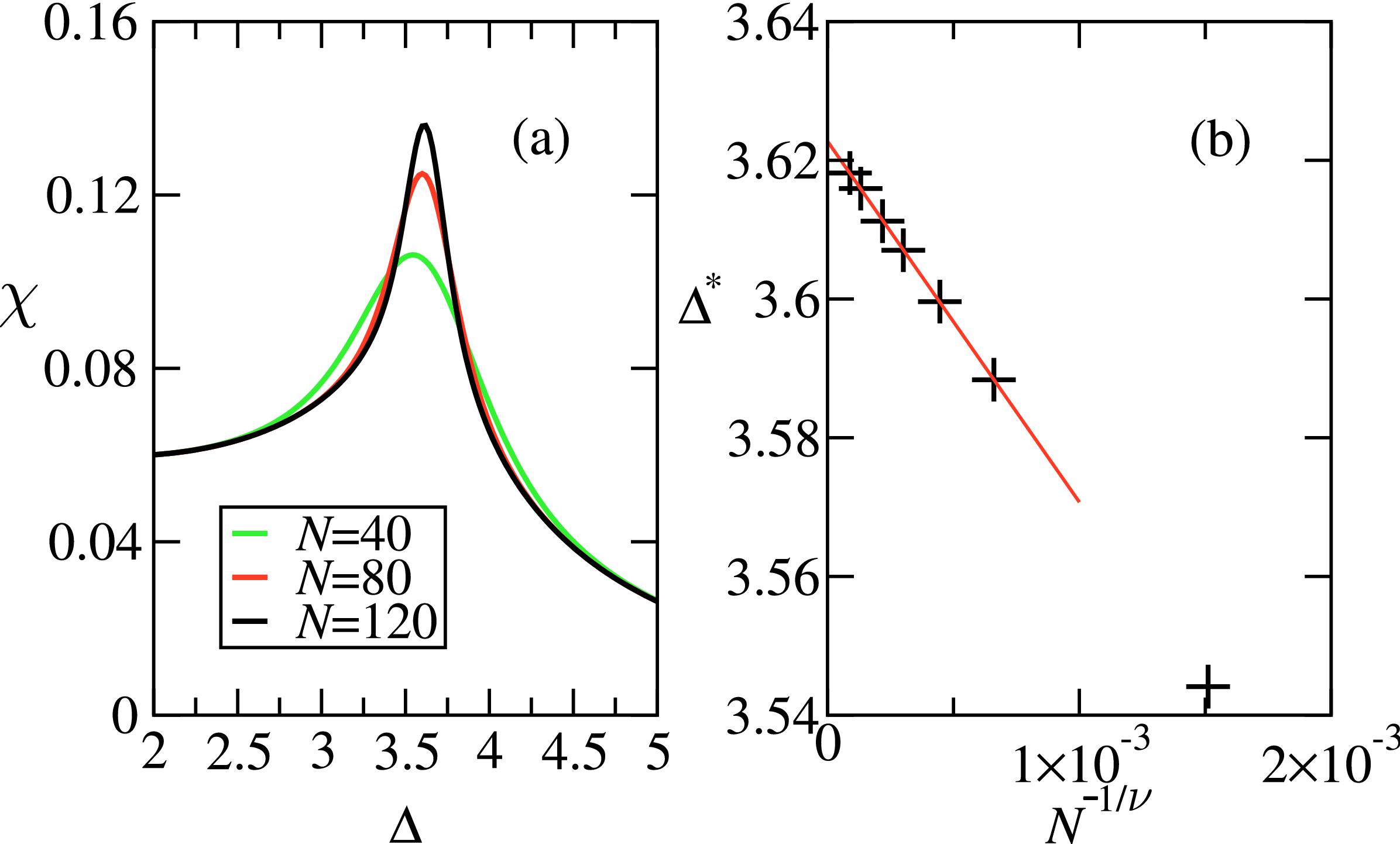}
\caption{(a) The 2nd derivatives of ground state energy density $\chi=-\partial^2 e_0/\partial\Delta^2$ as a function of $\Delta$ for fixed $\delta=0.3$ with different sizes $N$. (b) The extrapolation of the pseudo-critical points $\Delta^*$. It is obtained $\Delta_c\approx 3.623$ with the exponent $\nu\approx 0.568$.}
\label{fig:diff2e}
\end{figure}

According to Ehrenfest's classification of phase transitions, the $n$th order quantum phase transition presents non-analyticity of the $n$th derivatives of ground state energy density at the critical point. It has been firstly shown that the 2nd derivative of ground state energy diverges at the 2nd order quantum critical point, but it remains a finite value for the 3rd and 5th order quantum phase transitions.~\cite{Tzeng2008b} We report the results of energy derivatives for finding the critical points of the bond-alternating \textit{XXZ} model Eq.~(\ref{BA-XXZ-H}). The ground state energy per site $e_0=-\frac{1}{N}\sum_kR(k)$ for $\Delta=0$ can be exactly obtained from the previous subsection. In the thermodynamic limit $N\to\infty$, $e_0=-\frac{4}{\pi}I(1-\delta^2)$, where $I(x)=\int_0^{\frac{\pi}{2}}\sqrt{1-x\sin^2\theta}d\theta$ is the complete elliptic integral of the second kind. When $\delta=0$, we have $e_0=-\frac{\pi}{4}$, $\partial_\delta e_0=0$, and $\partial^2_\delta e_0\to-\infty$. Therefore, for the case $\Delta=0$, the quantum phase transition at $\delta=0$ belongs to second-order.

When $\Delta\neq 0$, by using DMRG, the 2nd derivative is calculated by the finite difference formula
\be\label{eq:2nd-finite}
\frac{\partial^2 e_0(g)}{\partial g^2}\approx\frac{e_0(g-\epsilon )-2e_0(g)+e_0(g+\epsilon)}{\epsilon^2}.
\ee
Where $e_0(g)$ is the ground state energy per site, $g$ is the parameter of the Hamiltonian Eq.~(\ref{BA-XXZ-H}), and $\epsilon$ is taken to be $5\times 10^{-3}$.

In absence of bond alternation, $\delta=0$, it is known the system undergoes a Berezinskii-Kosterlitz-Thouless (BKT) quantum phase transition at the critical point $\Delta_c=1$.~\cite{QP1D} The BKT quantum phase transition is an infinite-order transition, and the $n$th derivative of energy diverges only if $n\to\infty$. However, the order of the transitions could be different and depend on the path in the phase diagram.
It was firstly shown by Cross and Fisher~\cite{Cross1979} that the ground energy density of bond-alternating Heisenberg chain is in proportion to $\delta^{4/3}$, and the 2nd derivatives of energy density $\partial^2_\delta e_0\propto\delta^{-2/3}$. Thus the 2nd derivatives of energy density diverges and indicates a second order quantum phase transition at $\delta=0$.
On the other hand, for varying $\Delta$ and fixed $\delta=0.3$, as shown in Fig.~\ref{fig:diff2e}(a) and (b), the 2nd derivative of energy density diverges at the critical point $\Delta_c\approx 3.623$ in the thermodynamic limit $N\to\infty$. The values of the critical points obtained by the energy derivatives thus provide references for the values determined by the R\'enyi entropies $S_2$ and $S_\infty$. The precise phase diagram determined by the energy derivatives and the R\'enyi entropies is shown in the Fig.~\ref{fig:BAxxz}(b).


\footnotesize
\textbf{Acknowledgements.}
L.D. is grateful to MOST in Taiwan for the financial supports via MOST-104-2811-M-005-012.
Y.C.T. is grateful to Prof. Guang-Yin Chen and Prof. Wen-Min Huang for providing computing facilities. Y.C.T. is grateful to Hsing Tian Kong Culture \& Education Development Foundation for the scholarship.
L.C.K. acknowledges support from the National Research Foundation and Ministry of Education, Singapore.

\textbf{Author contributions statement.}
Y.C.T. \& L.D. initialized the project, analysed the data and wrote the manuscript. Y.C.T. developed the DMRG program and performed numerical computations. L.D. calculated the non-interacting fermions. MCC provided ref.~\onlinecite{CFM-Peschel} for the cfm formula. L.A. \& L.C.K. provided new perspective and interpretation, revised manuscript and supervised the project.


\end{document}